\documentclass[12pt, oneside]{article}   	
\usepackage{geometry}                		
\usepackage{graphicx}				
\usepackage{amsmath,amssymb,amsbsy,hyperref,multicol}

\title{On the capture rates of big bang neutrinos by nuclei\\
  within the Dirac and Majorana hypotheses}
\author{Esteban Roulet$^1$ and Francesco Vissani$^2$\\
\small $^1$  CONICET, Centro At\'{o}mico Bariloche, Argentina \\
\small $^2$ INFN, Laboratori Nazionali del Gran Sasso, Assergi (AQ), Italia}
\date{}							

\begin{document}
\maketitle
\renewcommand{\abstractname}{} 
\begin{abstract}
\vskip-7mm
\noindent The capture rates of non-relativistic neutrinos on beta decaying nuclei depends on whether their mass is Dirac or Majorana. 
It is known that for relic  neutrinos from the big-bang,
and within minimal assumptions, the rate is a factor two larger in the Majorana case. 
We show that this difference also depends on the value of the lightest neutrino mass and on the type of mass hierarchy. If the lightest neutrino has a mass below the meV, so that it is still relativistic today, its capture rate for the case of Dirac masses becomes equal to that for Majorana masses.
As a consequence, for the case of normal neutrino mass hierarchy, for which the total 
capture rate is dominated by the contribution from the lightest neutrino, if this one is below the meV  the distinction between the Dirac and Majorana scenarios  can only rely on the detection of the two heavier neutrinos, which is something very challenging.

\end{abstract}
\vskip1mm

\section{Introduction}
In the non-relativistic limit the 
Dirac and Majorana neutrinos behave differently, see e.g.~\cite{nr1}. This feature has practical  
interest for the observation of the neutrinos produced in the big-bang  that have been slowed down by the adiabatic expansion of the Universe. 
The process of neutrino capture by nuclei 
\begin{equation} \label{p0}
\nu_{\mbox{\tiny e}} + (A,Z) \to {\mbox{e}}^- + (A,Z+1) 
\end{equation}
does not have an energy threshold  for beta radioactive nuclei (it is `exothermic'). It thus offers, in principle,  a chance to probe the very low energy cosmological neutrinos~\cite{wei}.
The option of using a tritium target to detect the relic neutrinos is being considered seriously (see e.g. \cite{pt1,pt2}). In this connection, it was argued that, under minimal assumptions, the rate for  Majorana neutrinos is  a factor of two larger than for Dirac neutrinos \cite{luna}. This result is interesting {\em per se} and further motivates experimental efforts devoted to observe directly and for the first time the neutrinos produced in the big-bang.

In this note, we would like to quantify more precisely the difference in the interaction rate of Majorana and Dirac neutrinos in order to clarify that the factor two is just the maximum possible value. We  will argue that, 
in completely plausible particle physics scenarios~-~when the lightest neutrino mass is very light and the mass hierarchy is normal - 
the difference between the two can be much less, namely about~15\%.

\section{Capture rate of cosmic neutrinos}
Neutrinos of all three flavors were in equilibrium through weak interactions in the early Universe. They decoupled from the plasma at temperatures of about an MeV so that a population of relativistic left-chiral neutrinos and right-handed antineutrinos (the states participating in weak interactions) was present at that time with  a Fermi-Dirac thermal distribution. As the Universe cooled by the subsequent expansion their momenta got redshifted and the present distribution of this cosmic neutrino background (C$\nu$B), ignoring eventual small gravitational clustering effects to be discussed later, is
\begin{equation}
  f(p)=\frac{1}{1+\exp(p/T_\nu)},
\end{equation}
with $p\equiv|\vec{\rm p}|$ being the magnitude of the neutrino momentum and $T_\nu=1.945$~K being the present neutrino temperature parameter.\footnote{This temperature is smaller by a factor $(4/11)^{1/3}$ than the cosmic microwave background (CMB) temperature $T_\gamma =2.7$K, due to the photon reheating after $e^+e^-$ annihilation. There is actually also a very slight distortion  of the high energy tail of the neutrino distribution resulting from  $e^+e^-$ annihilations \cite{dolgov}, that we will neglect. Note that when neutrinos become non-relativistic, their energy distribution {\em is not} a thermal distribution with temperature~$T_\nu$.} After being produced, the neutrinos  quickly decohere into the three mass eigenstates $\nu_i$ with masses $m_i$. 
The initial left-chiral neutrinos, being relativistic, had left-handed helicity, $s_\nu=-1/2$ (recall that the helicity is the projection of the spin in the direction of motion) and, in the absence of interactions, this helicity remains conserved afterwards. Similarly, the right-chiral antineutrinos will appear at present as states with right-handed helicity, $s_\nu=+1/2$.  In the case of Majorana neutrino masses, there is no distinction between neutrinos and antineutrinos so that one can consider that there is a  density of left and right-handed helicity neutrinos $n_i(s_\nu)$, for each of the neutrino masses $m_i$, with $n_i(+1/2)=n_i(-1/2)\equiv n_\nu$, where $n_\nu=56$\,cm$^{-3}$ is the present average cosmic neutrino density. In the case of Dirac masses one has to distinguish between the densities of  neutrinos, $n_i$, and antineutrinos, $ \bar n_i$, so that one has at present $\bar{n}_i(+1/2)=n_i(-1/2)= n_\nu$, while  $\bar{n}_i(-1/2)=n_i(+1/2)=0$.

The interaction rate of these relic neutrinos with a tritium target with a number of atoms $N_{\mbox{\tiny T}}$ is \cite{cocco,luna}
\begin{equation} \label{rate}
\Gamma_{{\small {\rm C}\nu  {\rm B}}}=N_{\mbox{\tiny T}}\overline{\sigma}\,c\sum_{s_\nu=\pm1/2}\sum_{i=1}^3 |U_{ei}|^2\,n_i(s_\nu)C(s_\nu) ,
\end{equation}
with $\overline{\sigma}=3.834\times 10^{-45}$\,cm$^2$, $c$ is the speed of light  and
\begin{equation}
  C(s_\nu)=1-2s_\nu\beta_i,
\end{equation}
with $\beta_i\equiv v_i/c$ in terms of the neutrino velocity $v_i$, i.e. $\beta_i=p/\sqrt{p^2+m_i^2}$. The neutrino mixing factors $U_{ei}$ appear because each neutrino mass eigenstate $\nu_i$ couples to the capture process through its electronic component. The determination of the PMNS leptonic mixing parameters leads to $(|U_{e1}|^2,|U_{e2}|^2,|U_{e3}|^2)\simeq(0.68,0.30,0.022)$.

In the case of Majorana neutrinos both helicity components contribute to the rate, so that
\begin{equation} \label{gammamaj}
\Gamma_{\small  {\rm C}\nu {\rm  B}}^{\rm M}= N_{\mbox{\tiny T}}\overline{\sigma}\,c\sum_{i=1}^3 |U_{ei}|^2\,n_i\left[C(+1/2) +C(-1/2) \right]=2 N_{\mbox{\tiny T}}\overline{\sigma}\,c\sum_{i=1}^3 |U_{ei}|^2\,n_i.
\end{equation}
Moreover, if we can neglect the mass dependent gravitational clustering effects so that all neutrino densities equal $n_\nu$, using the unitarity of the PMNS mixing matrix one has that
\begin{equation} \label{gammamaj2}
\Gamma_{\small {\rm C}\nu{\rm B}}^{\rm M}= 2 N_{\mbox{\tiny T}}\overline{\sigma}\,c\,n_\nu.
\end{equation}
This gives, for a target of 100~g of tritium, a capture rate of about 8 events per year.

In the case of Dirac neutrino masses, due to the lepton number conservation the antineutrino states are unable to interact with the tritium nuclei to produce an electron from the conversion of a neutron into a proton, while the interaction with a proton ($\overline{\nu}_ep\to ne^+$) is kinematically forbidden.  Hence, only the left-handed helicity states interact, leading to 
\begin{equation} \label{gammadir}
\Gamma_{\small {\rm C}\nu{\rm B}}^{\rm D}= N_{\mbox{\tiny T}}\overline{\sigma}\,c\sum_{i=1}^3 |U_{ei}|^2\,n_iC(-1/2) \simeq N_{\mbox{\tiny T}}\,\overline{\sigma}\, c \,n_\nu(1+\sum_{i=1}^3 |U_{ei}|^2\langle\beta_i\rangle).
\end{equation}
The rate now involves the average of the neutrino velocity over the momentum distribution, i.e.
\begin{equation} \label{avbeta}
\langle\beta_i\rangle=\frac{\int_0^\infty \beta_i \,f(p)\,p^2\,{\rm d}p}{\int_0^\infty f(p)\,p^2\,{\rm d}p}.
\end{equation}
Let us note that if all three species of neutrinos are very slow, so that their velocities can be neglected, one just has $\Gamma^{\rm D}_{{\rm C}\nu{\rm B}}\simeq \Gamma^{\rm M}_{{\rm C}\nu{\rm B}}/2$, and hence a rate of only about 4 events per year in a target of 100~g of tritium would be expected in this limit if neutrinos are Dirac particles.

The average velocities of the different mass eigenstate neutrinos that appear in the rate in the Dirac case, Eq.~(\ref{gammadir}),  depend directly on the values of the neutrino masses. These are however unknown since the neutrino oscillation experiments only provide the mass square differences $\Delta m^2_{21}\equiv m_2^2-m_1^2\simeq 7.4\times 10^{-5}$~eV$^2$ and similarly $|\Delta m^2_{31}|\simeq 2.5\times 10^{-3}$~eV$^2$.
In terms of the lightest neutrino mass $m_\ell$, one has that for the normal hierarchy (NH, corresponding to $\Delta m^2_{31}>0$) and  inverse hierarchy (IH, corresponding to $\Delta m^2_{31}<0$)  cases 
the three neutrino masses will be given by \begin{eqnarray}
  (m_1,\,m_2,\,m_3)_{\rm  NH}&=&\left(m_\ell,\sqrt{\Delta m^2_{21}+m_\ell^2},\sqrt{\Delta m^2_{31}+m_\ell^2}\right)\nonumber \\
  (m_1,\,m_2,\,m_3)_{\rm IH}&=&\left(\sqrt{m_\ell^2-\Delta m^2_{31}},\sqrt{\Delta m^2_{21}+m_\ell^2-\Delta m^2_{31}},m_\ell\right).
  \end{eqnarray}

On the other hand, an upper bound  results from CMB and large scale structure observations for the sum of the three neutrino masses, $\Sigma\equiv \sum_im_i<230$~meV at 95\%~CL \cite{planck}. This bound, together with the measured mass squared differences, implies that the individual neutrino masses should all be smaller than 90~meV, with the lightest one\footnote{Simple expressions for the lightest neutrino mass as a function of $\Sigma$ are given in \cite{ss}.} 
being lighter than about 70~meV.
Some even stricter bounds have been obtained recently including new Lyman alpha data, where for instance in  \cite{yeche} the bound $\Sigma<140$~meV at 95\%~CL was derived.  This would imply that all neutrino masses are below about 65~meV, with the lightest one being lighter than about 40~meV. 
All this in principle allows for the lightest neutrino to be much lighter than the others, in which case its velocity may no longer be negligible at present and this could  modify the expected capture rates in the case of Dirac masses.

\begin{figure}[t]
 \centerline{ \includegraphics[width=8.cm]{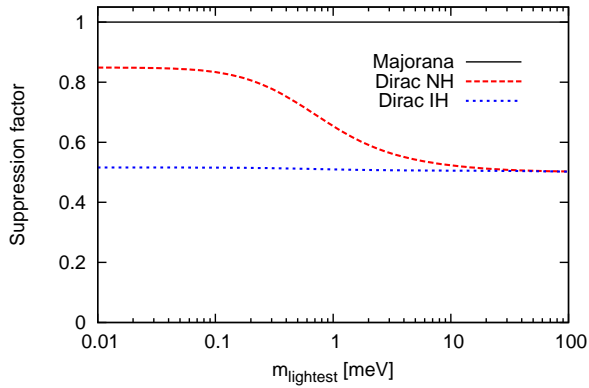}}
\caption{\footnotesize Numerical calculation of the suppression factor 
for the Dirac neutrino capture process, as a function of the lightest neutrino mass and for the two neutrino mass hierachies (normal and inverted).
For comparison, also the  case of Majorana neutrinos is shown, for which the suppression is 1 or, in other words,  
there is no suppression.\label{vg}}
\end{figure}

We may define the suppression factor $S$, that gives a measure of the reduction of the capture rate in the case of Dirac neutrinos, as
\begin{equation}
  S\equiv \frac{\Gamma^{\rm D}_{{\rm C}\nu{\rm B}}}{ \Gamma^{\rm M}_{{\rm C}\nu{\rm B}}}=\frac{1}{2}\left(1+\sum_{i=1}^3 |U_{ei}|^2\langle\beta_i\rangle\right).
\end{equation}
The resulting suppression is shown in Fig.~\ref{vg} as a function of the lightest neutrino mass, for both the normal and inverse hierarchies. Its value is $\sim 1/2$ if all neutrinos are non-relativistic but it becomes larger if some of the neutrino velocities, and in particular that of the lightest one, is non-negligible. Since for non-relativistic neutrinos the typical neutrino velocity is $\langle \beta_i\rangle \simeq  \langle p\rangle/m_i \simeq 3.15\, T_\nu/m_i\simeq 0.05(10$\,meV$/m_i)$, the effects of neutrino velocities on the rates can become non-negligible only if $m_\ell\ll 10$~meV. Note that even for $m_\ell=0$ one has that the second lightest neutrino would be heavier than $\sqrt{\Delta m^2_{21}}\simeq  8.6$~meV while the third one would be heavier than $\sqrt{|\Delta m^2_{31}|}\simeq  50$~meV. Hence, the changes in the suppression factor $S$  are  determined essentially by the contribution from the lightest neutrino alone.

The dependence of the capture rates upon the neutrino velocities, through the coefficients $C(s_\nu)$  in Eq.~(\ref{rate}), is due to a general feature of the charged-current neutrino interactions, namely the presence of the left-handed chirality projector. In fact, $C(-1/2)/2 = (1+\beta_i)/2$   is just
the probability that a left-handed helicity state  be in a left-handed chirality state.\footnote{Likewise,  $C(+1/2)/2 =(1-\beta_i)/2$  corresponds to the probability that a right-handed helicity state   be in a left-handed chirality state. While for Dirac neutrinos this state does not participate in the inverse decay process, being forbidden by the  lepton number,
in the case of Majorana neutrinos it does.}
In order to verify this statement, consider the plane wave solution for the left-handed helicity state, that in the Dirac representation reads 
\begin{equation}
  \Psi_{-}=
 \begin{pmatrix}\, \sqrt{1+m/E}\, \varphi_{-} \\[1ex] -\sqrt{1-m/E}\, \varphi_{-}  \end{pmatrix}
\frac{e^{-i(E  t-\vec{p}\cdot \vec{x})}}{\sqrt{2V}},
\end{equation}
where the bi-spinor is normalized as $\varphi_{-}^\dagger \varphi_{-}=1$ and $V$ is the formal volume of the space used for the quantization, see e.g., \cite{vanessa}. In the same representation the left chiral projector is 
$P_{\sf L}=\footnotesize\frac{1}{2}\! \begin{pmatrix}+1&-1\\-1&+1\end{pmatrix}$. One then finds that the left-chiral state $\Psi_{\sf L}=P_{\sf L}\Psi_{ -}$ has a  probability
\begin{equation}
  \int {\rm d}^3x \, \Psi_{\sf L}^\dagger\Psi_{\sf L}=\frac{1}{2}\left(1+\frac{p}{E}\right)=\frac{1}{2}\left(1+\beta\right).
  \end{equation}
One may say that as the velocity increases, the left-chiral component of the left-helicity neutrino state also increases and this explains why in the Dirac case the overall coupling of the neutrino to the inverse beta process becomes larger as $m_\ell \to 0$. In this limit, the lightest neutrino will be a left-chiral state, interacting with full strength, while the other two neutrinos are non-relativistic and interact less: thus,  one gets $S\to (1+|U_{e\ell}|^2)/2$. In the case of the NH, the lightest neutrino is $\nu_1$ and hence its capture rate is enhanced due to the large mixing $|U_{e1}|^2\simeq 0.68$, leading to $S^{\mbox{\tiny NH}}\to 0.84$. On the other hand, for the IH the lightest neutrino is $\nu_3$, and due to its small mixing $|U_{e3}|^2\simeq 0.02$ one has that the suppression factor is not significantly affected, with $S^{\mbox{\tiny IH}}\to 0.51$.

\section{On the detectability of the neutrino capture}
  As we have seen, in the case of Dirac neutrinos and when the lightest neutrino is sufficiently light, $m_\ell<{\rm meV}$, its capture rate on $\beta$-radioactive nuclei will be enhanced. 
  This effect is relevant for the normal hierarchy scenario since in this case the lightest state is $\nu_1$, for which the mixing to the beta process is largest,  $|U_{e1}|^2=0.68$. In this context it is important to note that there is now growing evidence from oscillation experiments favoring, at more than the 3$\sigma$ level, the NH scenario over the IH one \cite{nu2018}. We will hence mostly focus on the NH case in the following.

  The main difficulty that arises for the observation of the neutrino capture process  is that it involves electron energies higher by just about $2m_\nu$ with respect to the end-point of the ordinary $\beta$ decay of the target, and these decays provide an overwhelming rate (for instance, for 100~g of tritium there are about $10^{24}$ $\beta$ decays per year). Hence, in order that the signal from the C$\nu$B capture not be buried below the $\beta$ decay background, a very precise energy reconstruction is required. In particular, the full width half maximum (FWHM) energy resolution\footnote{The relation between the reconstructed and the true electron energies is assumed to be spread with a Gaussian with standard deviation $\sigma$, and the FWHM width is $\Delta\simeq 2.35\, \sigma$.} $\Delta$ needs to satisfy $\Delta<0.7\, m_\nu$ in order to be able to distinguish the signal arising from the  capture  of a neutrino with mass $m_\nu$ from the associated background from the $\beta$ decay~\cite{volpe,luna,blennow}. 

  The proposed PTOLEMY experiment \cite{pt1,pt2} aims at a resolution $\Delta\simeq 50$~meV and hence it may be sensitive to the capture of neutrinos with mass $m_\nu>70$~meV. In the case of NH, and given the cosmological constraints, this may only happen for the state $\nu_3$. Given its small coupling, $|U_{e3}|^2=0.022$, one may then expect that for 100~g of tritium a rate of only about one event per decade could result if neutrinos are Dirac (and twice as much in the Majorana case).  In the IH scenario both $\nu_1$ and $\nu_2$  may have masses above 70~meV so that the observable rates could  be as large as 4 per yr (or 8 per yr for the Majorana case).

  However, to explore neutrino masses lower than the meV, so as to be able to test the enhancement of the capture of Dirac neutrinos described before, correspondingly small energy resolutions would be required, something certainly challenging if at all possible.
For neutrino masses below the meV there is an additional effect that needs to be accounted for and that may be helpful, which is the fact that since the captured neutrino contributes all its energy to the electron energy, when the neutrino is relativistic this is more than just $m_\nu$. Hence, the electrons from the C$\nu$B capture will have a kinetic  energy  $K_e=K^0_{\rm end}+E_\nu$,  where $K^0_{\rm end}$ is the endpoint of the beta spectrum that would be obtained if neutrinos were  massless (the actual endpoint for the emission of the different neutrinos will be at $K_{\rm end}=K^0_{\rm end}-m_i$). Hence, even for $m_1\to 0$ the energy resolution required to resolve the peak of the electrons produced in the capture of $\nu_1$ would be finite and just slightly smaller than  $ \langle E_{\nu_1}\rangle\simeq 3.15\, T_\nu \simeq 0.5$~meV. 
  
\begin{figure}[t]
\centerline{\includegraphics[width=7.cm]{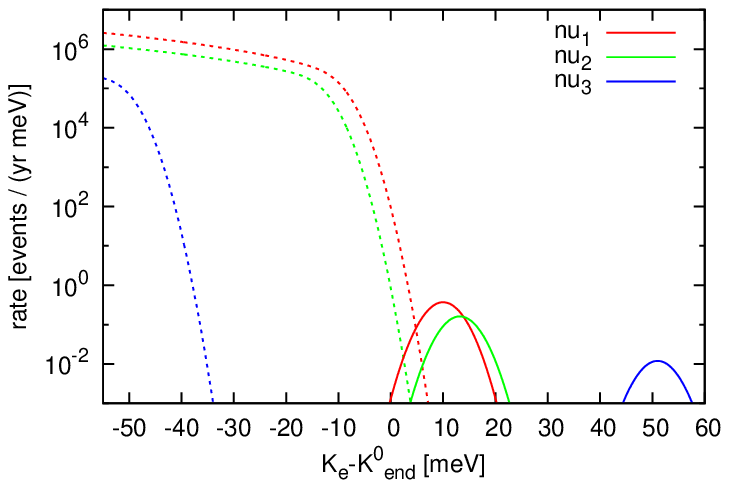}\includegraphics[width=7.cm]{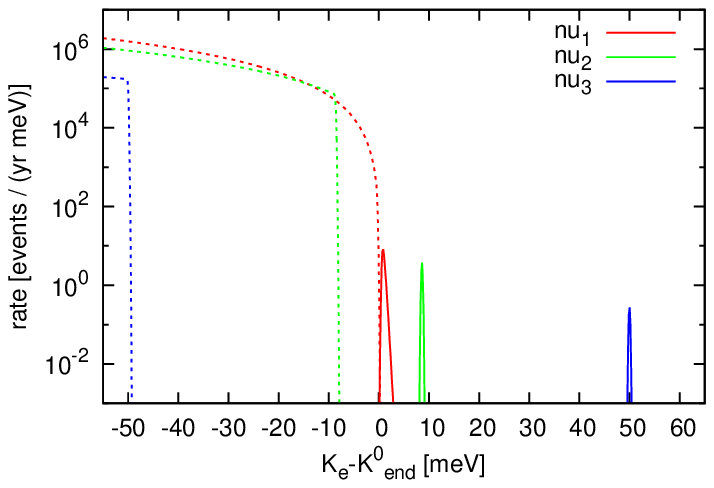}}
\caption{\footnotesize Rate of beta decay (dashed lines) and C$\nu$B capture  for Dirac neutrinos (solid lines) and for a target with 100~g of tritium. We assume the normal hierarchy  and show separately the contributions associated to each of the  three neutrino mass eigenstates for two different futuristic scenarios for the detection. The left panel adopts the lightest neutrino mass as $m_1=10$~meV and the energy resolution $\Delta=7$~meV. The right panel has instead $m_1=0.3$~meV and $\Delta=0.3$~meV. \label{rates.fig}}
\end{figure}

In Fig.~\ref{rates.fig} we show the differential rate of produced electrons, as a function of the electron kinetic energy, for a tritium target of 100~g.
Dashed lines correspond to the $\beta$ decays and solid lines to the C$\nu$B capture. 
The separate contributions from the three neutrino mass eigenstates, assumed to be Dirac neutrinos, are shown. 
The left panel adopts the lightest neutrino mass as $m_\ell=10$~meV and an energy resolution $\Delta=7$~meV. The right panel is for $m_\ell=0.3$~meV and $\Delta=0.3$~meV. Several of the main features just discussed can be understood examining these two situations.

In the scenario considered on the left panel, the energy resolution $\Delta\simeq 0.7\, m_1$ is just enough to separate the peak resulting  from the capture of the lightest neutrino from the large  background from the  $\beta$ decays. The peaks from the three neutrinos can hence be observed, with $\nu_1$ contributing about 68\% of the rate, $\nu_2$ contributing about 30\% and $\nu_3$  about 2\%. In view of the above discussion, for Dirac neutrinos one would hence expect on average about three events per year from $\nu_1$ and one from $\nu_2$.

The panel on the right of Fig.~\ref{rates.fig} describes an even more optimistic situation for what concerns the detector performances, namely,
 a much smaller energy resolution, that may hopefully be obtained  one day  in the future. 
 We illustrate there the expected differential rates for $m_1=0.3$~meV and $\Delta=0.3$~meV. One can see that the signal from $\nu_1$ extends in this case up to about 1~meV and is wider than the peaks from the other neutrinos, due to the non-negligible kinetic energies of the $\nu_1$s. For Dirac neutrinos, the capture rate associated to $\nu_1$ will be almost the same as that in the case of Majorana neutrinos (about 6 events per year) while for $\nu_2$ and $\nu_3$ the rates are half those in the Majorana case, making the total rate from the three neutrinos to be about 7 events per year, corresponding to about 85\% of the total rate of the Majorana case. Note that this result will hold as long as $\Delta<0.3$~meV and no matter how small is the $\nu_1$ mass.

\begin{figure}[t]
\centerline{\includegraphics[width=7.cm]{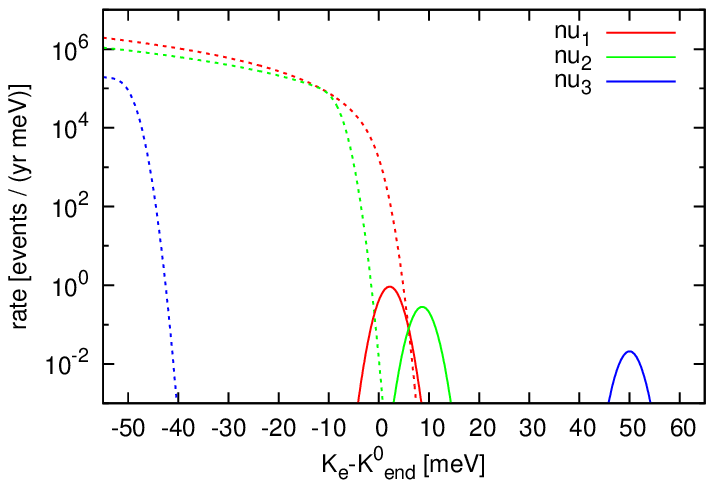}\includegraphics[width=7.cm]{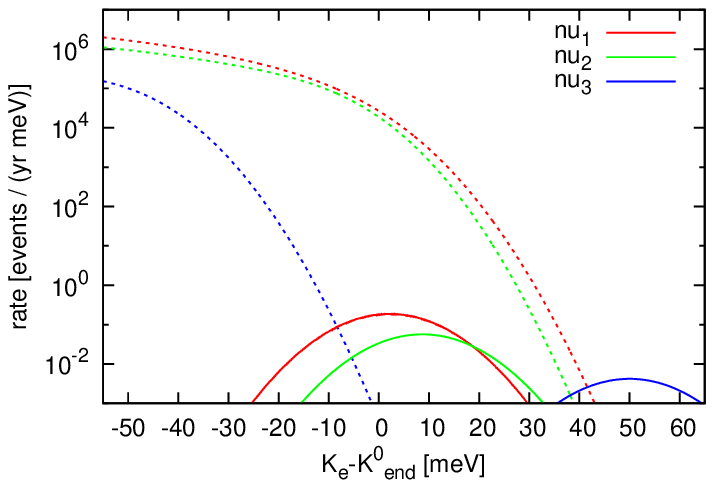}}
\caption{\footnotesize Rate of beta decay (dashed lines) and C$\nu$B capture  for Dirac neutrinos (solid lines) and for a target with 100~g of tritium. We assume the normal hierarchy  and show separately the contributions associated to each of the  three neutrino mass eigenstates, adopting  $m_1=1$~meV. The left panel considers an energy resolution $\Delta=4$~meV. The right panel has instead $\Delta=20$~meV. \label{nu23.fig}}
\end{figure}

On the other hand,  the peak from $\nu_2$ may still allow to discriminate between the Dirac and Majorana scenarios, although this would require target masses larger than a few hundred grams of tritium so as to have several events per year. This would be feasible even if the resolution were larger, as long at it is smaller than $\sim 4$~meV, so as to be able to resolve the  peak from $\nu_2$.
Note that in the case of very small $m_1$, to separate the signal of the second neutrino $\nu_2$ from the background of the beta decay associated to the lightest neutrino $\nu_1$, which has an enhanced rate proportional to $|U_{e1}|^2$ and a larger end point than the beta decay signal from $\nu_2$, requires that $\Delta<0.5\,m_2\simeq 4$~meV. Similarly, to separate the signal of $\nu_3$ from the background associated to $\nu_1$ would  require that $\Delta<0.4\,m_3\simeq 20$~meV. Hence, with energy resolutions between 4 and 20~meV one may no longer identify the signal from $\nu_2$ but may  still identify the signal from the heavier neutrino $\nu_3$. However, to be able to gather a reasonable number of events from it, so as to discriminate between Dirac and Majorana scenarios, would require target masses of few kilograms of tritium. These last two cases are illustrated in Fig.~\ref{nu23.fig}, where we show the capture rates (and beta decay backgrounds) associated to the three neutrinos, adopting NH and $m_1=1$~meV, for energy resolutions of $\Delta=4$~meV (left plot) and $\Delta=20$~meV (right plot).

\section{Effects of gravitational clustering}
Due to the gravitational attraction by structures like galaxies and clusters of galaxies, which are actually dominated by the dark matter, massive neutrinos will have a tendency to concentrate towards their deep gravitational potential wells. This effect is largest the more massive are the structures and the larger are the neutrino masses. The neutrinos that may end up being trapped inside those structures are those of the low momentum tail of the distribution of the relic neutrinos, typically those that have velocities below the value $v_{\rm esc}$ that is required to escape from the central parts of the structures \cite{gelmini}. These velocities may be of a few thousand km/s for the largest clusters, that have masses $M\sim 10^{15}M_\odot$, while they can be of order of 500~km/s for galaxies like the Milky Way, having masses  $M\sim 10^{12}M_\odot$. Note that the fraction of the cosmic neutrinos lying in this low momentum tail,  $F(v<v_{\rm esc})$,  is small
\begin{equation}
  F(v<v_{\rm esc})=\frac{\int_0^{m_iv_{\rm esc}}  \,f(p)\,p^2\,{\rm d}p}{\int_0^\infty f(p)\,p^2\,{\rm d}p}<1\%,
\end{equation}
where the upper bound indicated is that corresponding to a neutrino mass of 50~meV  and we adopted $v_{\rm esc}=500$~km/s. One hence expects that most of the relic neutrinos will remain unclustered and having a distribution which will only be slightly affected by the gravitational effects. 
Several detailed studies of the gravitational clustering of massive neutrinos around dark matter and baryonic structures have been performed (see e.g. \cite{ringwald, zalda, pastor, zhang}), and the conclusion is that for the cosmologically allowed neutrino masses the local neutrino density enhancement at the Earth location is not expected to be large, just of the order of 10\% for $m_\nu=50$~meV. The local density of clustered neutrinos $n^c_i$ can be parameterized as $n_i^c\simeq \rho^c(m_i)n_\nu$, where  an approximate analytical expression for the enhancement factor $\rho^c$, obtained for neutrino masses in the range $40<m_\nu/{\rm meV}<150$, is \cite{zhang}
\begin{equation}
\rho^c(m_\nu)\simeq 1+0.1(m_\nu/50~{\rm meV})^{2.21}.
 \end{equation}
Accounting for this gravitational clustering,  the rate of capture of neutrinos in a tritium experiment would be
\begin{equation} \label{p0}
\Gamma_{\small {\rm C}\nu{\rm B}}=N_{\mbox{\tiny T}}\overline{\sigma}\,c\sum_{i=1}^3 |U_{ei}|^2\,\left[ n_i(-1/2)(1+\langle \beta_i\rangle) +  n_i(+1/2)(1-\langle \beta_i\rangle)\right]\rho^c(m_i).
\end{equation}
Let us give a brief very qualitative discussion of the different effects that could result from the clustering of the neutrinos in the Milky Way (a more detailed evaluation of these effects is anyhow not necessary since for the masses considered the effects turn out to be quite small).
An unclustered  neutrino from the cosmic background should gain a velocity of order $v_{\rm esc}$ when falling down to the Earth location.  Since the typical background neutrino velocities are much larger, this should affect $\langle \beta_i\rangle$, and hence the capture rates,  by a negligible amount. This should also not affect the neutrino helicity in any significant way. On the other hand, the low velocity tail of the local neutrino distribution could get enhanced by the contribution from the neutrinos that got trapped during the formation and evolution of the dark matter halos and galaxies.  These neutrinos will be non-relativistic and, moreover, since their momentum will constantly change as they orbit the structure, an equal amount of left and right-handed helicity components will develop. In first approximation one could then assume that near the Earth there will be a density $n_\nu$  of unclustered left-handed helicity neutrinos, having the original distribution $f(p)$, together with a density $(\rho^c-1)n_\nu$ of clustered neutrinos, having  low velocities ($v<v_{\rm esc}$) and having equal amounts of left and right-handed helicities.  An analogous reasoning can be applied to the right-handed antineutrino helicity states.  The discussion we made in the previous Section would hence apply directly to the  unclustered component. The new contribution from the clustered neutrinos, which are non-relativistic and have equal amounts of both helicity components, should give an additional contribution to the capture rates, which for the Dirac case is
\begin{eqnarray}
  \delta( \Gamma_{\small {\rm C}\nu{\rm B}}^{\rm D})^c  &\simeq &N_{\mbox{\tiny T}}\overline{\sigma}\,cn_\nu \sum_{i=1}^3 |U_{ei}|^2\,\left[\rho^c(m_i)-1\right]  \frac{C(-1/2)+C(+1/2)}{2} \nonumber \\
  &=&N_{\mbox{\tiny T}}\overline{\sigma}\,c n_\nu \sum_{i=1}^3 |U_{ei}|^2\,\left[\rho^c(m_i)-1\right].
  \end{eqnarray}
For the Majorana case the contribution will be twice as large.
We see that, given that $|U_{ei}|^2[\rho^c(m_i)-1]\ll 1$, specially for the NH case in which the state with the largest value of $|U_{ei}|^2$ will be the one with the smallest over-density, the impact of the clustered contribution is expected to be negligible for the presently allowed neutrino masses.

\section{Conclusions}

The detection of the background of cosmic neutrinos left over from the big bang would certainly be a major milestone for physics and cosmology. An additional bonus from this detection could be the possibility to distinguish whether neutrinos are Dirac or Majorana particles, something which otherwise may only be achievable through  the searches of neutrinoless double beta decays. This possibility results from the fact that the relic neutrinos, which at present are left-handed helicity neutrinos and right-handed helicity  antineutrinos, can only be captured through their left-handed neutrino chirality component. Since the cosmic neutrinos are usually expected to be non-relativistic at present, they will have equal amounts of both chirality components but, while in the Dirac case the antineutrinos will not be captured by the target, in the Majorana case they will (since there is no distinction between neutrinos and antineutrinos in this last case).

However, we have shown that if the lightest neutrino is sufficiently light, $m_\ell\ll 10$~meV, it will still be dominantly left-chiral today and hence it will interact with full strength (while the antineutrinos will be right-chiral and hence will not interact even in the Majorana case). Hence, the capture rate of the lightest neutrino will not be sensitive  to whether neutrinos
are Dirac or Majorana particles if $m_\ell<1$~meV. This result is more relevant for the NH case, for which the lightest state is $\nu_1$, having the largest coupling to the neutrino capture process. The possibility to discriminate between the Dirac and Majorana  scenarios should then rely in this case on a detailed measurement of the rate associated to the second neutrino $\nu_2$, which in the NH has a mass $m_2\simeq 8.6$~meV in the limit $m_1\to 0$.  This would however require very small energy resolutions, $\Delta<4$~meV, and tritium target masses of few hundred grams so as to have sufficiently large rates. With  larger energy resolutions one may still identify the signal from the heavier neutrino $\nu_3$, as long as $\Delta<20$~meV, but to have sizeable rates would require target masses of few kilograms.

\section*{Acknowledgments}
E.R. acknowledges support from ANPCyT (grant PICT 2016-0660) and CONICET (grant PIP 2015-0369).
F.V. is grateful for the invitation at Instituto Balseiro, Bariloche (Argentina) within the {\em Programa Maldacena de Profesores Invitados.}


\end{document}